\documentclass[review]{elsarticle}
\usepackage{fancyhdr}
\usepackage{multicol} 
\usepackage{comment}
\usepackage{amsmath}
\numberwithin{equation}{section}
\usepackage{amssymb}
\usepackage{natbib}
\usepackage{graphicx} 
\usepackage{booktabs}

\allowdisplaybreaks
\newcommand{\smallMatrix}[1]{\begin{pmatrix}#1\end{pmatrix}}

\begin{document}

\begin{frontmatter}

\title{Investigation of crackling noise in the vibration isolation systems of the KAGRA gravitational wave detector}

\author[mymainaddress,nomal,naoj]{Yuhang Zhao\corref{mycorrespondingauthor}}
\cortext[mycorrespondingauthor]{Corresponding author}
\ead{yuhang@icrr.u-tokyo.ac.jp}
\author[mymainaddress]{Shin Kirii}
\author[mymainaddress]{Yingtao Liu}
\author[mymainaddress]{Takashi Uchiyama}
\author[naoj]{Ryutaro Takahashi}
\author[mymainaddress,nu]{Seiji Kawamura}

\address[mymainaddress]{KAGRA Observatory, ICRR, The University of Tokyo, Hida, Gifu 506-1205, Japan}
\address[nomal]{Department of Astronomy, Beijing Normal University, Beijing 100875, China}
\address[naoj]{Gravitational Wave Science Project, National Astronomical Observatory of Japan, Mitaka. Tokyo 181-8588, Japan}
\address[nu]{Division of Particle and Astrophysical Sciences, Nagoya University, Nagoya, Aichi, 464-8602, Japan}

\begin{abstract}
Gravitational wave detectors, such as KAGRA, require complex mirror suspension systems to reach high sensitivity. One particular concern in these suspensions is the presence of crackling noise, where motion of the steel crystal structure disturbs material defects, causing impulse force release and exciting mirror motion. Crackling in the Geometric Anti Spring filters (GAS) in KAGRA's suspension system could introduce noise in the gravitational wave detection band. We investigate crackling noise in a miniature GAS with a low-frequency cyclical stress. We developed a scaling law between miniature and KAGRA GAS. Applying experimental results to the scaling law allows us to estimate the upper limit of crackling noise in KAGRA. It is found that crackling noise should be below the target sensitivity for gravitational waves above 60\,Hz.
\end{abstract}


\end{frontmatter}

\section{Introduction}
The detection of gravitational waves provides a unique and new way to observe the gravitational universe. In 2015, a gravitational wave \cite{1} was detected for the first time by the two US detectors comprising the Laser Interferometer Gravitational-wave Observatory (LIGO) \cite{2_ligo}. Soon after, Advanced Virgo \cite{3_virgo} joined observation of LIGO. This simultaneous observation brought a significant breakthrough for multi-messenger astronomy, the detection of GW170817 \cite{4_170817}. In the observing period of the third round, GW190521 \cite{5_190521} was detected with an intriguing feature of its corresponding source masses located in the pair-instability mass gap. Up to now, two gravitational-wave source catalogs have been released \cite{6_cata, 7_cata}, including information of 50 gravitational wave signals. The Japanese underground cryogenic detector KAGRA \cite{8_aso} has also  finished its first observing run and is undergoing an upgrade period. The operation of KAGRA will not only improve the localization of gravitational wave signals \cite{9_prospects} but also bridge the gap between the current and the next generation detectors \cite{10_kagra}.

\begin{figure}[htbp]
\begin{center}
\includegraphics[clip, width=7cm]{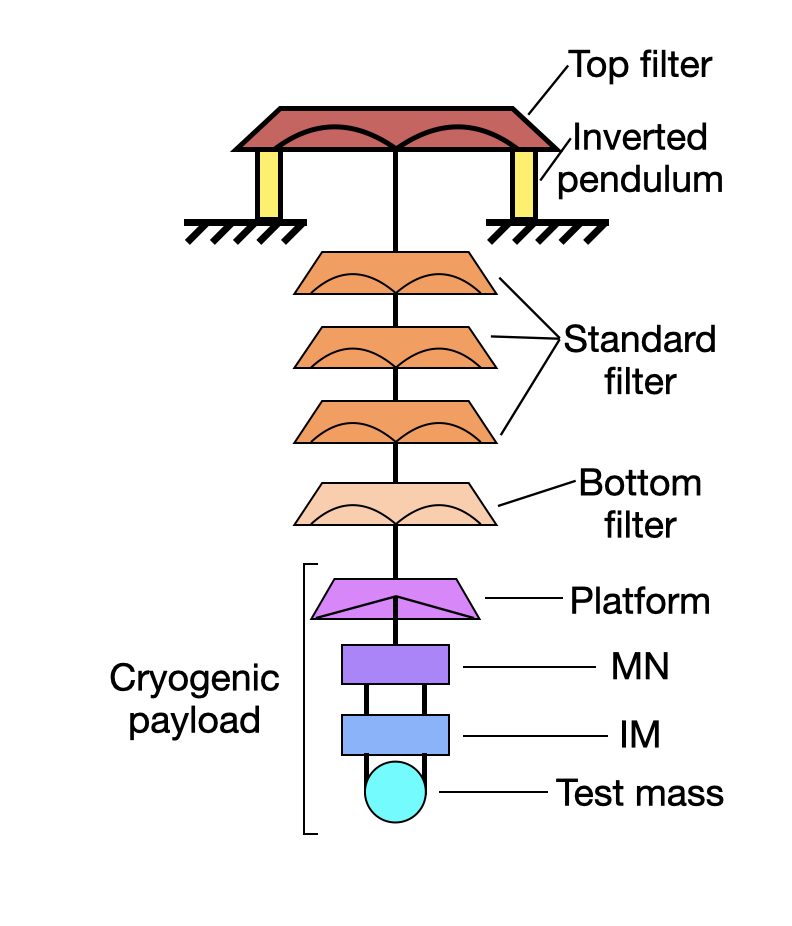}
\caption{Schematic of type-A suspension system for the four test-mass mirrors used in KAGRA. Wire length is not to scale and much longer in reality. For the last three stages, each has its recoil masses but not shown in this schematic. This figure is modified from the figure in~\cite{michimura} to highlight the platform stage is not GAS. 'MN' stands for marionette. 'IM' stands for intermediate mass. }\label{typeat}
\end{center}
\end{figure}

\begin{figure}[htbp]
\begin{center}
\includegraphics[clip, width=6cm]{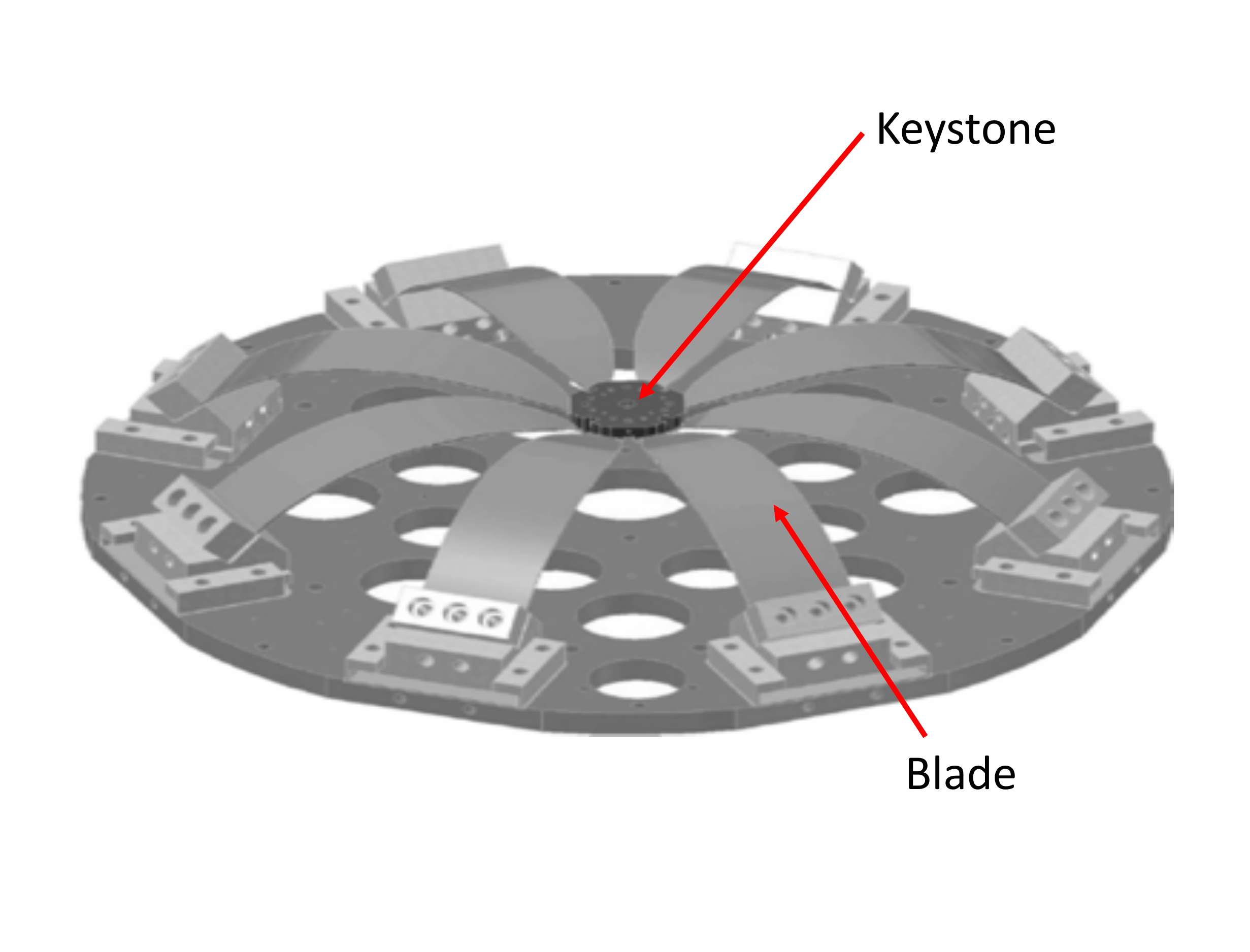}
\caption{General type of the GAS used in the suspension system of KAGRA. The diameter of the disk frame is, for example, 0.7\,m for the bottom filter. The wide side of each blade is fixed on the common disk. The narrow side of each blade is fixed on a common plate, called a keystone. The lower side of the keystone is loaded with heavy mass, which makes blades curve. The horizontal compression and load can be adjusted to have an optimum GAS resonant frequency value.}\label{gas}
\end{center}
\end{figure}

KAGRA is constructed underground inside the Kamioka mine, which helps to reduce seismic noise. But due to the strict requirement of mirror motion, like all GW detectors \cite{11_vajente}, KAGRA uses vibration isolation systems (VIS) that can attenuate seismic noise above their resonant frequencies of the suspended pendulums. In addition, based on different suspended mirrors' displacement noise contribution, different mirror suspension systems are used, such as type-A (shown in figure~\ref{typeat}) to achieve the highest vibration isolation for test masses \cite{13_hirose} and lower vibration isolation is realized by type-B \cite{21_akutsu}, type-Bp \cite{12_akiyama} or type-C. In each VIS, horizontal vibration isolation is provided by pendulum, while vertical vibration isolation is mainly provided by Geometric Anti-Spring filters (GAS) (shown in figure~\ref{gas}). One particular concern with GAS is crackling noise.

\begin{figure}[htbp]
\begin{center}
\includegraphics[clip, width=9cm]{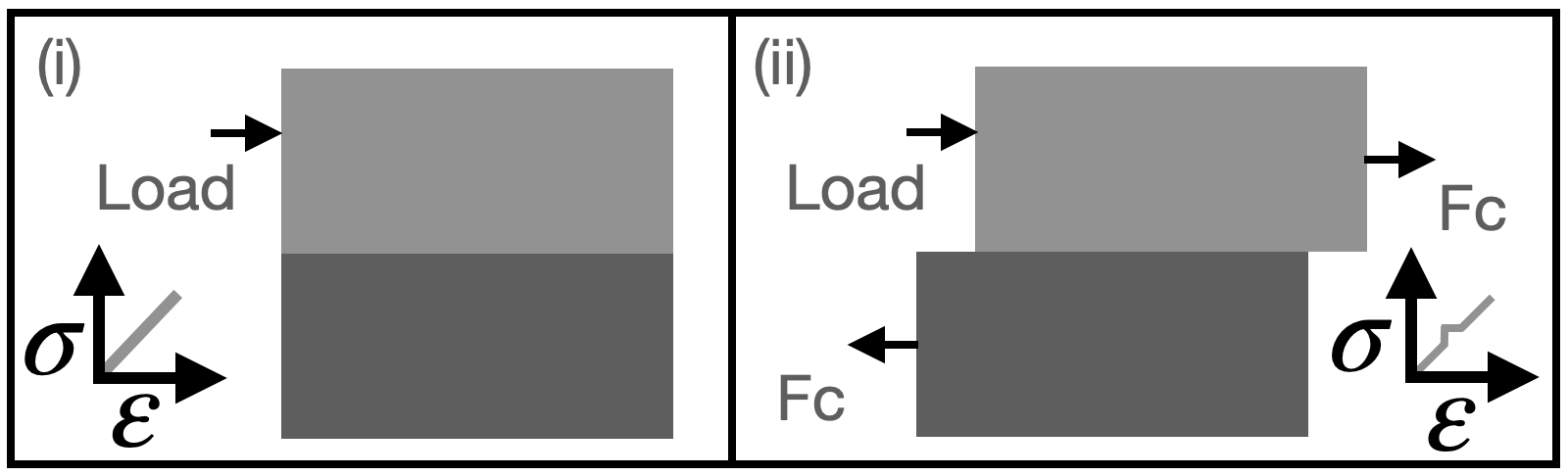}
\caption{Schematic representation of the crackling noise model. An oscillating load is applied to the system. $\sigma$ is the stress induced by the load. $\varepsilon$ is the strain induced by the stress. (i) stress and strain have linear relation as Hooke's law; no crackling happens. (ii) Crackling occurs when there is step-like relation between stress and strain. At the same time, a pair of impulsive force $F_c$ releases and acts on the different sides of dislocation.}\label{simple}
\end{center}
\end{figure}

Crackling noise is an unsteady and impulsive phenomenon that the internal strain in solid changes discontinuously when stress is externally applied \cite{15_sethna}. Crackling is known to occur in various materials, such as crystalline \cite{16_dimiduk}, amorphous \cite{15_sethna}, and granular \cite{granular} materials. The step-like stress response of material from strain was found to deviate from Hooke's law and cause sudden force release \cite{09_dahmen}. This sudden force release excites the motion of surrounding objects. A single crackling generation process can be represented in a simple model, as shown in figure~\ref{simple}. In gravitational wave detectors, such as KAGRA, passive filters in VIS work with the principle of spring to isolate seismic noise. Simultaneously, the low-frequency microseismic motion is not filtered but amplified by the resonance of the suspension system. Therefore, the suspension resonance applies an oscillating load to its corresponding stage of the suspension system, which creates step-like stress release and disturbs the test mass. We call a single step-like stress release a crackling event. These crackling events are so small that each of them is uncorrelated. In the end, the test mass of type-A is displaced by the incoherent sum of many crackling events. We call noise generated in this mechanism 'crackling noise' of gravitational wave detectors. Especially, a recent micro-scale experiment found step-like stress response exists even well below the material yielding point \cite{17_ni}. This is exactly the situation of materials in VIS, which makes the examination of crackling noise even more necessary.

Incidentally, creep noise is similar to crackling noise; both of them are originated from the dislocation movement. But the creep noise of a mirror suspension system is considered to be caused by large and stationary stress \cite{19_crack}, which comes from the relaxation properties of a mechanical device. The dislocation movement can be stopped if a pinning point is encountered. Afterwards, the de-pinned dislocation drives the creep noise and manifests as acoustic emission \cite{20_crack} or material extension \cite{21_crack}. In our research, we consider a noise caused by varying stress, which is a collective number of small dislocation movements. In particular, the crackling noise is driven by varying stress much smaller than yielding points. It should be noted that creep noise is out of the scope of this research.

In type-A, as shown in figure~\ref{typeat}, there are passive vertical damping filters between each stage. Among all the passive damping filters in type-A, GAS \cite{14_anti} has an enormous volume and is compressed with considerable stress, as shown in figure~\ref{gas}. So we expect it to be the primary source of crackling noise. However, the crackling noise originating from the upper stage of GAS will be attenuated by the lower stage of GAS. In type-A, the last stage of GAS is the ``bottom filter" (BF), which has the most significant impact on detector sensitivity and is the target of this research. Although the platform stage is the next downstream stage of BF (see figure~\ref{typeat}), the blades in the platform are not compressed in the horizontal direction, the platform has a lower temperature, and the number of blades is three in the platform instead of five in BF. Therefore, we assume the crackling noise from the platform is much smaller than BF.

In this paper, we discuss the impact of GAS crackling noise in KAGRA. To detect the crackling noise directly in the GAS configuration,  we measured two identical miniature GAS differential motions based on the Michelson interferometer principle. Since we performed this measurement on a miniature GAS, a scaling law was developed so that the measurement result could be scaled to the crackling noise in KAGRA. The experimental setup's sensitivity still needs to be improved, but this work is already enough to set an upper limit for crackling noise in KAGRA.

Incidentally an independent experiment \cite{18_vajente}, aiming for the characterization of crackling noise in LIGO's vertical vibration isolation single blade cantilever, was designed and carried out at the California Institute of Technology. According to the characteristic configuration of the LIGO suspension system, a detailed model \cite{19_vajente} was developed to calculate how crackling noise scales with the cantilever's size and geometry.

\section{Objectives and strategies}
The purpose of this investigation is to estimate the effect of crackling noise that could occur in KAGRA. For that purpose, we employed the following strategy. Firstly, we design a crackling-noise measurement system (CNMS) consisting of a specially-arranged GAS. This GAS is smaller than the one used in KAGRA. But it has the same design principle and material as KAGRA GAS. Then, we incorporate two identical GAS units inside an interferometer as two end mirrors. This way, we aim for direct detection of the uncorrelated crackling noise from these two identical GAS units. Then we convert the crackling noise (or an upper limit) obtained in the CNMS into the estimated KAGRA's crackling noise using a conversion ratio. The difference between CNMS and KAGRA are the following:\\
(1) \textbf{Different measured directions of mirror displacement}: The crackling noise that happens inside GAS causes displacement in the vertical direction. In CNMS, we aim for direct detection in the vertical direction. However, in KAGRA, crackling noise influences detector sensitivity due to vertical to horizontal coupling. \\
(2) \textbf{Different transfer functions from crackling induced force to mirror displacement}: The mirror is suspended as a single pendulum from the keystone of the GAS, and the disk frame of the GAS is rigidly attached to the frame structure of the suspension system in the CNMS. On the other hand, in KAGRA, the test-mass mirror is suspended with several stages of VIS from the keystone of the BF's GAS, while the disk frame of the BF's GAS is also suspended with several stages of VIS from the frame structure of the suspension system.\\
(3) \textbf{Different crackling event occurrence frequencies and magnitudes}: We expect the number of crackling events in a certain period of time will be different depending on the blade material and size. The event magnitude will be also different depending on material, excitation frequency, and excitation strength.

All the above differences are discussed in detail in the rest of this paper, which makes a conservative scaling of the crackling noise from CNMS to KAGRA. 

\section{Experimental setup}\label{setup}

\begin{figure}[htbp]
\begin{center}
\includegraphics[clip, width=8cm]{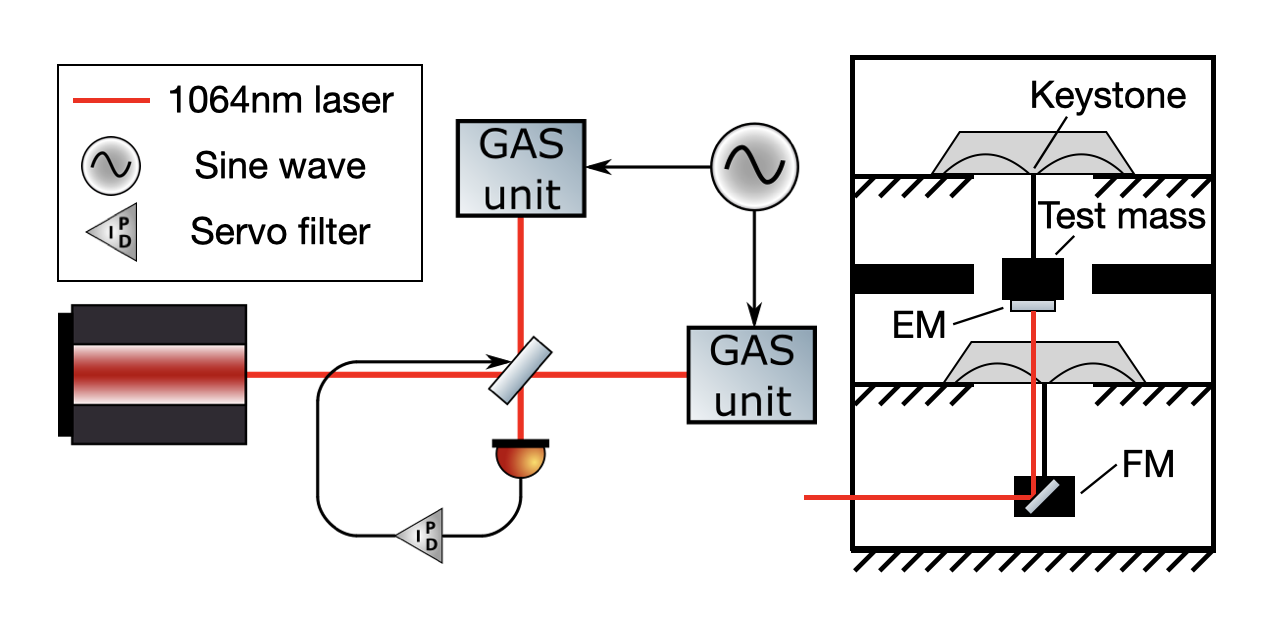}
\caption{Left: Schematic diagram of the CNMS for measuring crackling noise (top view). Right: Schematic diagram of the GAS unit (side view). The CNMS is a modified Michelson interferometer. Modification is to use a folding mirror (FM) to fold the laser direction to hit the end mirror (EM) vertically. This modification aims to detect differential keystone motion directly since crackling noise makes the keystone move in the vertical direction. The feedback signal to lock the interferometer to the mid-fringe is applied not to the EMs but to the BS to avoid disturbing the EM positions. When system noise is low enough, crackling noise should appear in the feedback signal to the BS. The keystones of both GAS are equipped with coil-magnet actuators to excite crackling noise.}\label{souti}
\end{center}
\end{figure}

In the CNMS, we try to detect crackling noise using a Michelson interferometer to remove common-mode noises. Figure~\ref{souti} is the scheme of the experimental setup. When the interferometer is operated around a half fringe, its power response $\delta P$ to mirror motion $\delta L$ is
\begin{eqnarray}
\delta P= P_0\frac{2\pi}{\lambda}\delta L,
\end{eqnarray}
where $P_0$ is the interferometer output power when operated at a bright fringe. This response makes CNMS sensitive to the differential motion of mirrors. Each arm of the Michelson interferometer has a "GAS unit". It is a combination of an End Mirror (EM) and a Folding Mirror (FM) for folding the horizontal beam to the vertical direction. The EM is attached to the lower part of the test mass, which is suspended by wire from the keystone of the GAS. The mass of the test mass is chosen to have a GAS resonant frequency around 3\,Hz. The FM is also suspended in the same manner independently under the EM suspension. The suspension of EM and FM provides seismic isolation in the vertical direction as $1/\mathrm{f^2}$ above 3Hz. The beam splitter (BS) is suspended as a double pendulum and provides seismic isolation in the horizontal direction as $1/\mathrm{f^4}$ above 3\,Hz. The interferometer is controlled to be maintained at a half fringe with a bandwidth around 1\,kHz. The error signal is taken from the photodetector and sent to the coil-magnet actuator of the BS. Large sinusoidal force is applied to the two keystones simultaneously by magnet-coil actuators to excite the crackling noise. It should be noted that since the lengths of the two arms change simultaneously, the correction signal (output of servo filters) does not contain this large signal at all in principle. However, the crackling noise, which occurs independently in the two arms, could appear in the correction signal. We calibrate the correction signal $x_c$ to mirror motion $x_2$ as
\begin{equation}
x_2 = \frac{1+H}{I\times F} x_c ,
\end{equation}
where $H$ is the open-loop transfer function, $I$ is the transfer function of the interferometer, and $F$ is the transfer function of the servo filters.

\section{Scaling crackling noise from CNMS to KAGRA type-A}\label{am}
To scale the crackling noise or its upper limit to the KAGRA sensitivity, as introduced, we have three aspects to be considered. In this section, we describe details about the three aspects of this scaling law. Incidentally it should be noted that there is another aspect to be considered: the attenuation effect of crackling noise inside blades. The attenuation of crackling noise inside GAS could be larger in KAGRA than CNMS since the noise needs to propagate a longer distance. This aspect is not considered in the scaling law here, because we are not confident in the scaling law of the attenuation effect. Once considered, this attenuation effect could make the estimation of crackling noise in KAGRA type-A smaller. Therefore, the scaling law introduced here can be considered a conservative one.

\subsection{Measurement direction of mirror displacement}
\begin{figure}[htbp]
\begin{center}
\includegraphics[clip, width= 6cm]{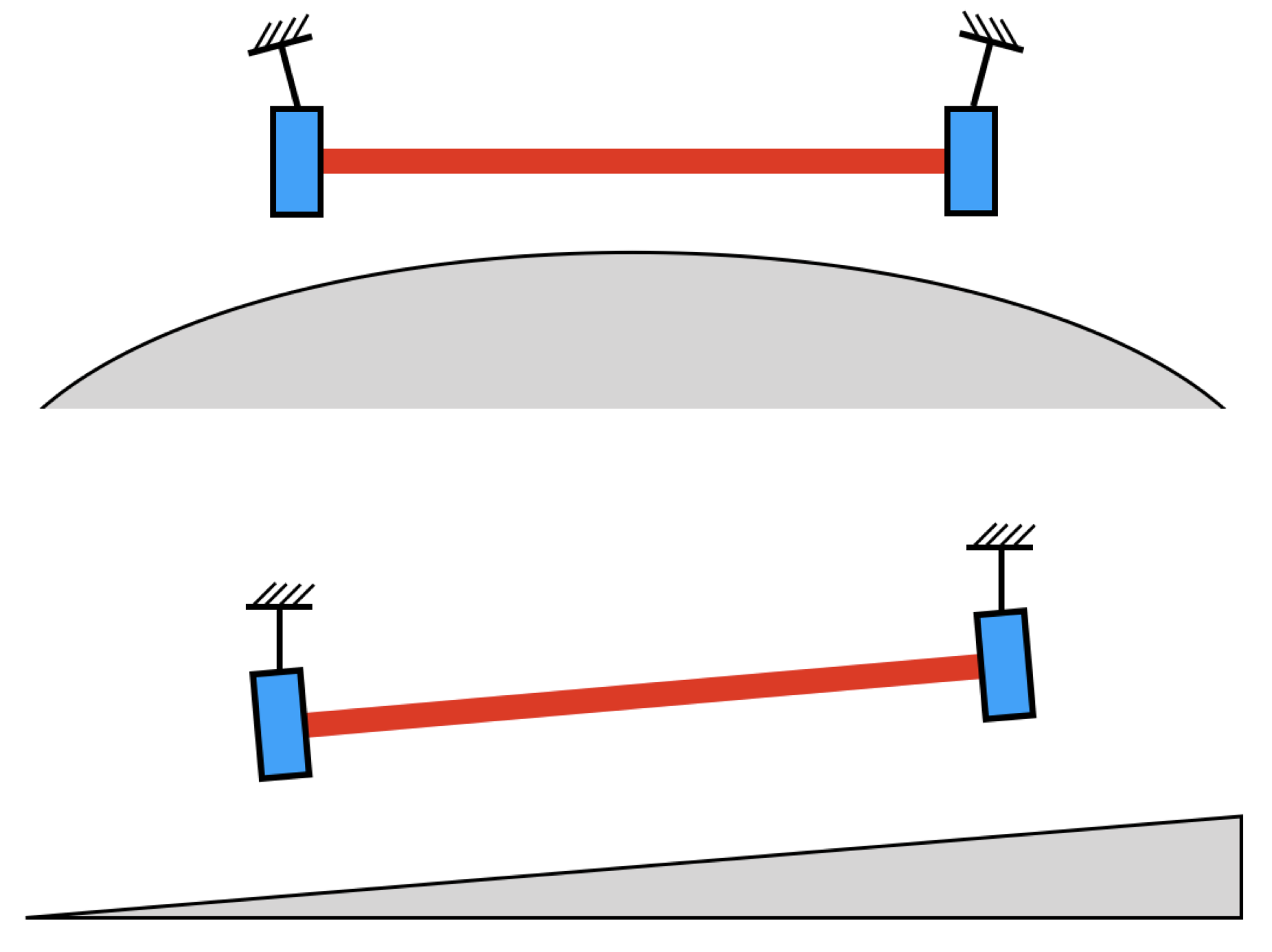}
\caption{(Upper) Mirror tilt caused by the curvature of the Earth. (Lower) Mirror tilt caused by the ground tilt. (The tilt angles are exaggerated in the figures.)}\label{coup}
\end{center}
\end{figure}

Crackling noise of GAS causes displacement noise in the vertical direction. In CNMS, the modified Michelson interferometer directly detects this vertical displacement noise. In KAGRA, even though we nominally measure the horizontal displacement of test masses, figure~\ref{coup} shows that there is horizontal-to-vertical coupling due to the curvature of the Earth and ground tilt. The ground tilt of the KAGRA arm tunnel is 1/300 ( = 3.33$\times 10 ^ {-3}$), and the difference between the optical axis and the horizontal axis due to the curvature of the Earth is $0.24\times 10^{-3}$. Therefore, in the worst case, the coupling from vertical displacement noise to horizontal direction is $C_{vtoh} = 3.57\times 10^{-3}$.

\subsection{Transfer function from crackling induced force to mirror displacement}
\begin{figure}[htbp]
\begin{center}
\includegraphics[clip, width= 8cm]{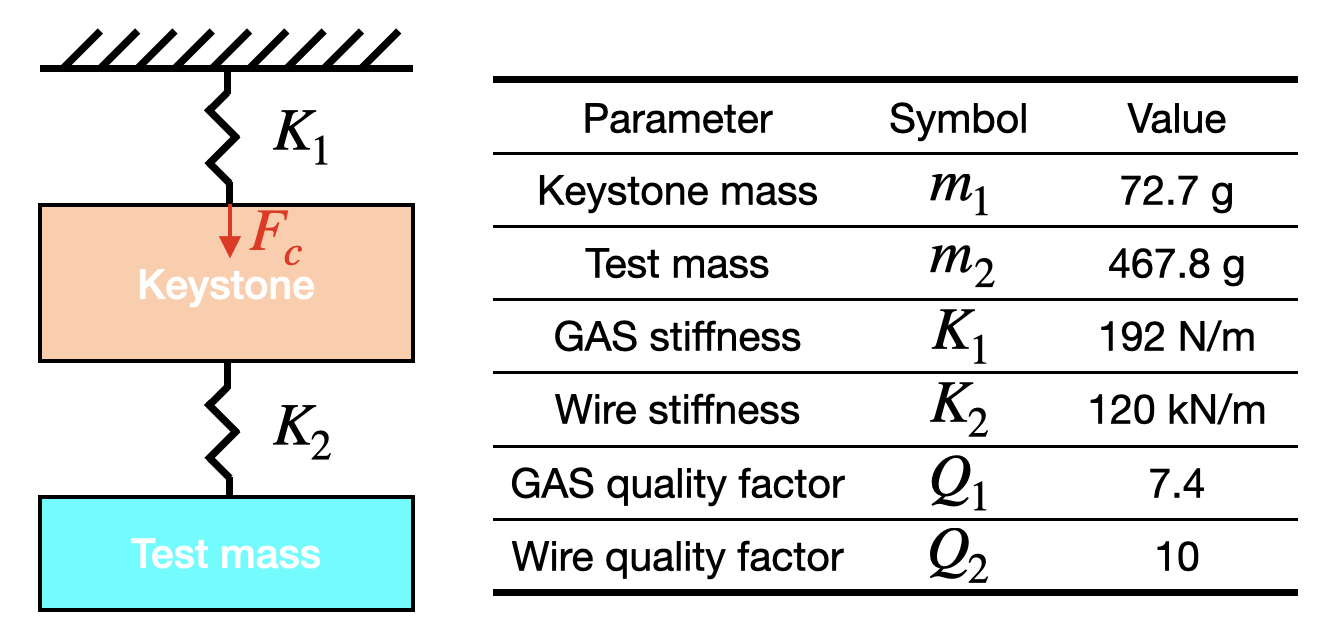}
\caption{Left: Simplified model of CNMS. Right: Parameters of CNMS. One side of GAS is fixed to the ground. The other side of GAS is fixed to the keystone of GAS. Crackling induced force $F_c$ is applied to the keystone. Another force with the same amplitude but the opposite direction is also applied to the ground, which is not shown since the ground is fixed. }\label{scheme_cnms}
\end{center}
\end{figure}

Both CNMS and KAGRA type-A are multi-degree-of-freedom systems. To get the response of the whole suspension system to crackling induced force, we construct matrixes for them separately as the method introduced in\cite{5}. Each element of the matrix corresponds to each stage of the suspension system. The equation of motion of the whole system can be expressed in general as
\begin{eqnarray}\label{eom}
[m]\ddot{\vec x} + [c]\dot{\vec x} +[k] \vec x = \vec f,
\end{eqnarray}
where $[m]$ is the mass matrix, $[c]$ is the damping coefficient matrix, $[k]$ is the stiffness coefficient matrix, $\vec x$ is the displacement vector, and $\vec f$ is the force vector. By taking the Laplace transform of equation~\ref{eom} and substituting $i\omega$ for $s$, we can express it as $-[m]\omega^2\vec x + [K]\vec x = \vec f$. We call $[K]$ generalized stiffness coefficient, which has a form of $K = k+i\omega\frac{\sqrt{mk}}{Q}$. Here, $Q$ is the quality factor of equivalent spring, which comes from $c = \sqrt{mk}/Q$. We describe the CNMS response to crackling induced force $F_c$ as shown in figure~\ref{scheme_cnms}.

\begin{figure}[htbp]
\begin{center}
\includegraphics[clip, width=8cm]{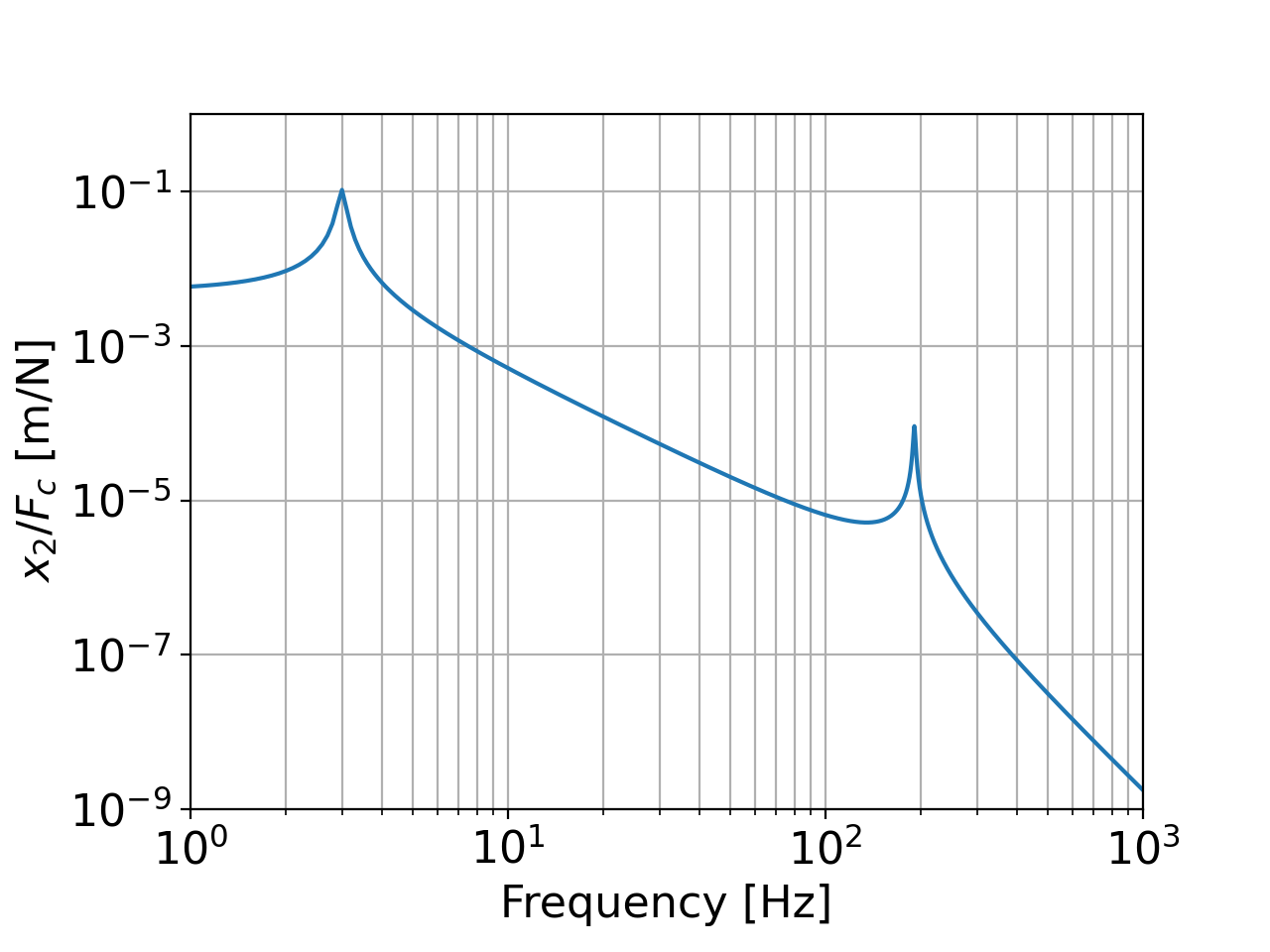}
\caption{Transfer function from crackling induced force $F_c$ to the displacement of the test mass $x_2$.}\label{tf_cnms}
\end{center}
\end{figure}

According to figure~\ref{scheme_cnms}, we can obtain the following equation of motion:
\begin{eqnarray}
\smallMatrix{-m_1\omega^2+K_1+K_2 & -K_2 \\ -K_2 & -m_2\omega^2+K_2}\smallMatrix{x_1 \\ x_2} = \smallMatrix{F_c \\ 0}.
\end{eqnarray}
The mirror displacement $x_2$ caused by the crackling induced force $F_c$ can be derived from the above equation. We obtain the transfer function as shown in figure~\ref{tf_cnms}.

\begin{figure}[htbp]
\begin{center}
\includegraphics[clip, width=14cm]{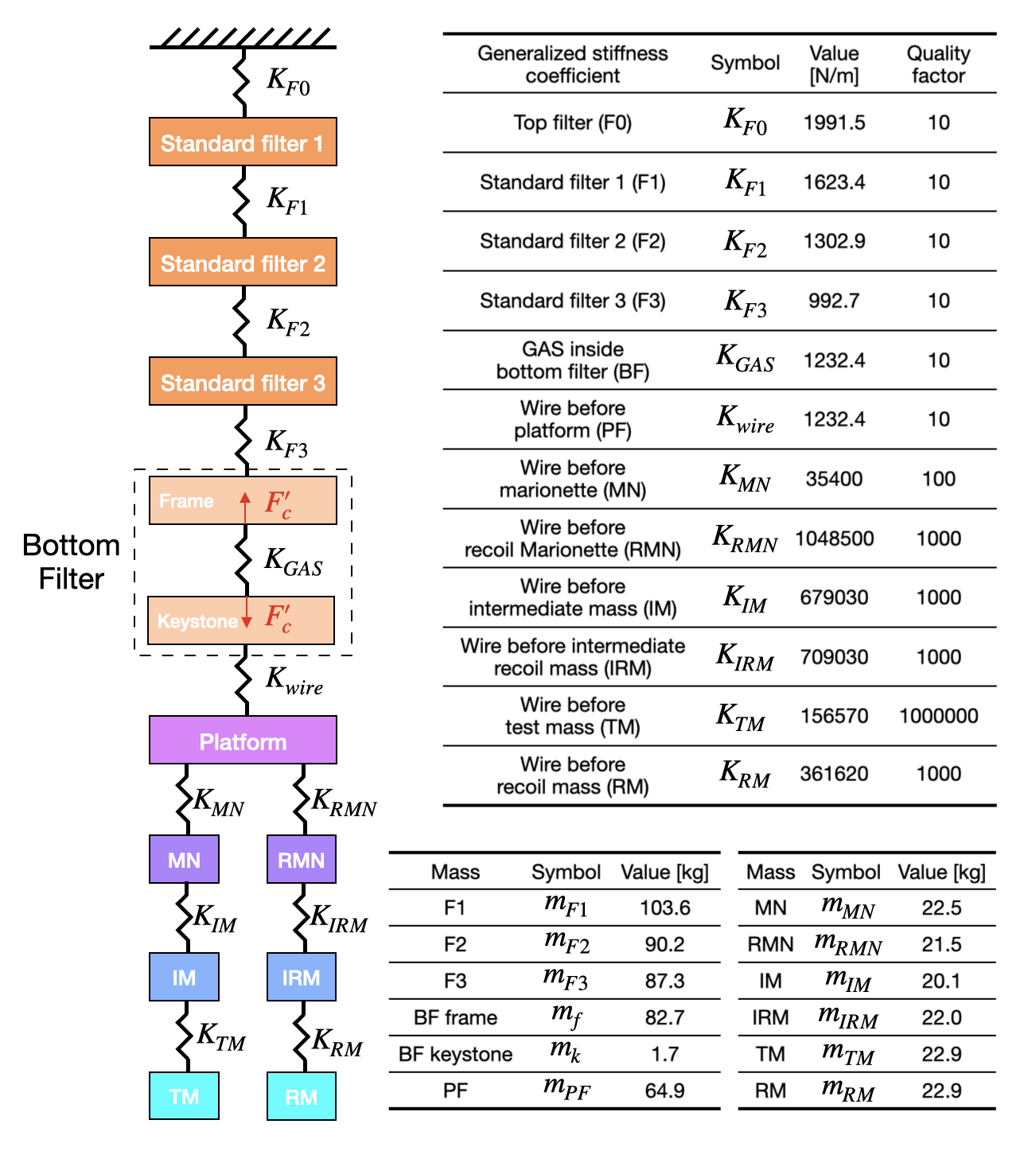}
\caption{Left: Scheme of crackling noise inside KAGRA type-A. Right: Parameters of type-A. Crackling noise happens inside the GAS of the bottom filter, which is indicated by a pair of force $F_c'$.}\label{type-a-eom}
\end{center}
\end{figure}

Figure \ref{type-a-eom} shows the scheme of crackling noise inside KAGRA type-A. The transfer function from crackling induced force $F_c'$ (happened inside BF) to the displacement of test mass x is calculated by solving equation \ref{eqn:typea}. 
\begin{eqnarray}\label{eqn:typea}
M_{typea} \vec x = \vec F,
\end{eqnarray}
where $M_{typea}$ is the matrix containing information of inertia force and stress, which is described in equation \ref{marix_typea}. $\vec x$ is the vector containing the information of displacement. $\vec F$ is the external force vector, which has non-zero term only from crackling induced force $F_c$.
\begin{eqnarray}\label{marix_typea}
M_{typea} = \smallMatrix{
    A1 & A2 & 0 & 0 & 0 & 0 & 0 & 0 & 0 & 0 & 0 & 0  \\
    B1  & B2 & B3  & 0 & 0 & 0 & 0 & 0 & 0 & 0 & 0 & 0 \\
    0 & C1  & C2  & C3  & 0 & 0 & 0 & 0 & 0 & 0 & 0 & 0 \\
    0 & 0 & D1  & D2 & D3 & 0 & 0 & 0 & 0 & 0 & 0 & 0 \\
    0 & 0 & 0 & E1 & E2 & E3 & 0 & 0 & 0 & 0 & 0 & 0 \\
    0 & 0 & 0 & 0 & F1 & F2 & F3 & 0 & 0 & F4 & 0 & 0 \\
    0 & 0 & 0 & 0 & 0 & G1 & G2 & G3 & 0 & 0 & 0 & 0 \\
    0 & 0 & 0 & 0 & 0 & 0 & H1 & H2 & H3 & 0 & 0 & 0 \\
    0 & 0 & 0 & 0 & 0 & 0 & 0 & I1 & I2 & 0 & 0 & 0 \\
    0 & 0 & 0 & 0 & 0 & J1 & 0 & 0 & 0 & J2 & J3 & 0 \\
    0 & 0 & 0 & 0 & 0 & 0 & 0 & 0 & 0 & K1 & K2 & K3 \\
    0 & 0 & 0 & 0 & 0 & 0 & 0 & 0 & 0 & 0 & L1 & L2
 } .
\end{eqnarray}
The elements in $M_{typea}$ are listed in equation \ref{elements}.
\begin{eqnarray}\label{elements}
A1 &=& -m_{F1}w^2 + K_{F0} + K_{F1}, \nonumber\\
A2 &=& -K_{F1}, \nonumber\\
B1 &=& -K_{F1}, \nonumber\\
B2 &=& -m_{F2}w^2 + K_{F1}  + K_{F2}, \nonumber \\
B3 &=& -K_{F2}, \nonumber\\
C1 &=& -K_{F2}, \nonumber\\
C2 &=& -m_{F3}w^2 + K_{F2}  + K_{F3}, \nonumber\\
C3 &=& -K_{F3}, \nonumber\\
D1 &=& -K_{F3}, \nonumber\\
D2 &=& -m_{f}w^2 + K_{F3}  + K_{GAS}, \nonumber\\
D3 &=& -K_{GAS}, \nonumber\\
E1 &=& -K_{GAS}, \nonumber\\
E2 &=& -m_{k}w^2 + K_{GAS} + K_{wire}, \nonumber\\
E3 &=& -K_{wire}, \nonumber\\
F1 &=& -K_{wire}, \nonumber\\
F2 &=& -m_{PF}w^2 + K_{wire} + K_{MN} + K_{RMN}, \nonumber\\
F3 &=& -K_{MN}, \nonumber\\
F4 &=& -K_{RMN}, \\
G1 &=& -K_{MN}, \nonumber\\
G2 &=& -m_{MN}w^2 + K_{MN} + K_{IM}, \nonumber\\
G3 &=& -K_{IM}, \nonumber\\
H1 &=& -K_{IM}, \nonumber\\
H2 &=& -m_{IM}w^2 + K_{IM} + K_{TM}, \nonumber\\
H3 &=& -K_{TM}, \nonumber\\
I1 &=& -K_{TM}, \nonumber\\
I2 &=& -m_{TM}w^2 + K_{TM}, \nonumber\\
J1 &=& -K_{RMN}, \nonumber\\
J2 &=& -m_{RMN}w^2 + K_{RMN} + K_{IRM}, \nonumber\\
J3 &=& -K_{IRM}, \nonumber\\
K1 &=& -K_{IRM}, \nonumber\\
K2 &= & -m_{IRM}w^2 + K_{IRM} + K_{RM}, \nonumber\\
K3 &=& -K_{RM}, \nonumber\\
L1 &=& -K_{RM}, \nonumber\\
L2 &=& -m_{RM}w^2 + K_{RM}. \nonumber
\end{eqnarray}

\begin{figure}[htbp]
\begin{center}
\includegraphics[clip, width=8cm]{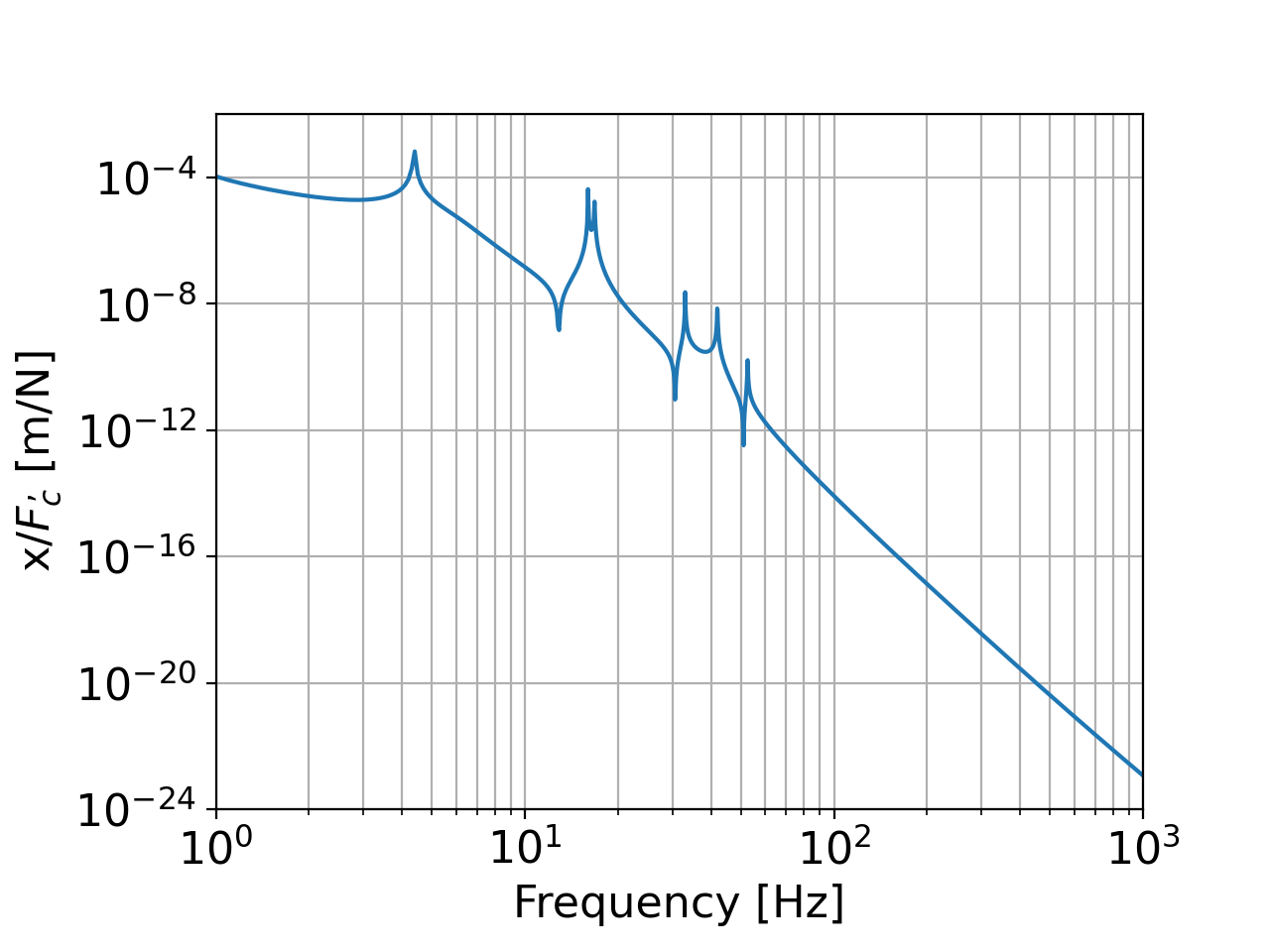}
\caption{Transfer function from crackling induced force $F_c'$ to the displacement of the test mass $x$.}\label{tf_typea}
\end{center}
\end{figure}

 Solving equation \ref{eqn:typea}, we obtained figure~\ref{tf_typea}, which is the transfer function from crackling induced force to the displacement of the test mass.

\subsection{Crackling noise occurrence frequency and magnitude}
Crackling noise is the incoherent sum of a large number of crackling events. Due to many differences between CNMS and KAGRA type-A, the crackling noise occurrence frequency and magnitude will be different. To obtain a reasonable noise scaling from CNMS to KAGRA type-A, we need to understand their difference. The number of crackling events needs to be considered in a certain period of time; we call it the crackling events occurrence frequency. Based on the model of crackling in figure~\ref{simple}, crackling events occurrence frequency should depend on the following factors
\begin{enumerate}
\item The material of the GAS blade. Different materials have a different amount of movable dislocations. For Maraging steel, its intermetallic precipitates trap the dislocation movement\cite{maraging_desalvo}. This makes Maraging steel to be used widely in the suspension system of gravitational wave detectors. The intermetallic precipitates are similar to the disorder in Ising model \cite{sethna}, which tends to stop the propagation of avalanches. Therefore, a larger amount of intermetallic precipitates may lead to a smaller cut-off avalanche size in the power-law event number distribution of avalanche size \cite{15_sethna}. In terms of crackling noise magnitude, different material exhibits different amplitude of micro-plasticity in small scale \cite{17_ni} or acoustic emission in large scale \cite{Faran}. Since we use the same material for both CNMS and BF of KAGRA type-A, the same avalanche size probability distribution and crackling noise magnitude should hold. 
\item The volume of the GAS blade. Blade volume decides the number of material defects and thus the number of crackling events.
\item The frequency of the oscillating load. For material close to the yielding point, the increasing load frequency increases crackling noise\cite{white}. However, the microplasticity of small-scale face-centered-cubic crystals shows more deviation from elasticity at lower load frequency \cite{17_ni}. These two types of research investigated the crackling noise coming from different response mechanisms to load frequency. Unfortunately, however, up to now, we are not certain about the exact relation between crackling noise and load frequency. Therefore, we assume crackling noise is proportional to $f^{\varepsilon}$ with $-1<\varepsilon<1$.
\item The amplitude of strain change caused by oscillating load. The stress release of crackling comes from the strain change. A larger strain change means more crackling events.
\end{enumerate}
The first two items come from material intrinsic properties, while the last two items come from the oscillating external load. In CNMS, load is provided as a sinusoidal wave. But in KAGRA type-A, load is the unavoidable and considerable micro-seismic motion. Considering material property and oscillating load, we find the collection of crackling noise
\begin{eqnarray}
N =  f^{\varepsilon} V  a_c \varrho,
\end{eqnarray}
where $V$ is the volume of the GAS blade, $a_c$ is the amplitude of strain change caused by oscillating load, and $\varrho$ is a proportionality constant, depending on the number density of the crystal imperfections and dislocations ($\varrho$ is common for the same material). 

\subsection{Scaling law}
Considering the three factors of the scaling law, we arrive at a final scaling factor for the displacement noise measured in CNMS as follows
\begin{eqnarray}
S(f) = \frac{(1/TF_{type-A})N_{type-A}}{(1/TF_{CNMS})N_{CNMS}}C_{vtoh}/L_{arm}, \label{convertfunction}
\end{eqnarray}
where $L_{arm}$ is the arm length of KAGRA, which is used to convert the displacement to the strain. Other parameters are provided in Table~\ref{tab: sca}.

\begin{table}[htb]
    \caption{Symbols and values for scaling law parameters.}
    \label{tab: sca}
    \centering
        \begin{tabular}{c c c}
        \toprule
        Parameter & Symbol & Value \\
        \midrule
		CNMS blade volume & $V_{CNMS}$ & 525 $mm^3$\\
		type-A blade volume & $V_{type-A}$ & 61380 $mm^3$\\
		coupling to horizontal & $C_{vtoh}$ & $3.57\times 10^{-3}$\\
		KAGRA arm length & $L_{arm}$ & 3000 m \\
		BF resonance amplitude & $\delta\gamma_{BF}$ & $5.7\times 10^{-8}$ m \\
        \bottomrule
        \end{tabular}
\end{table}

\section{Experiment and results}
\begin{figure}[htbp]
\begin{center}
\includegraphics[clip, width=8cm]{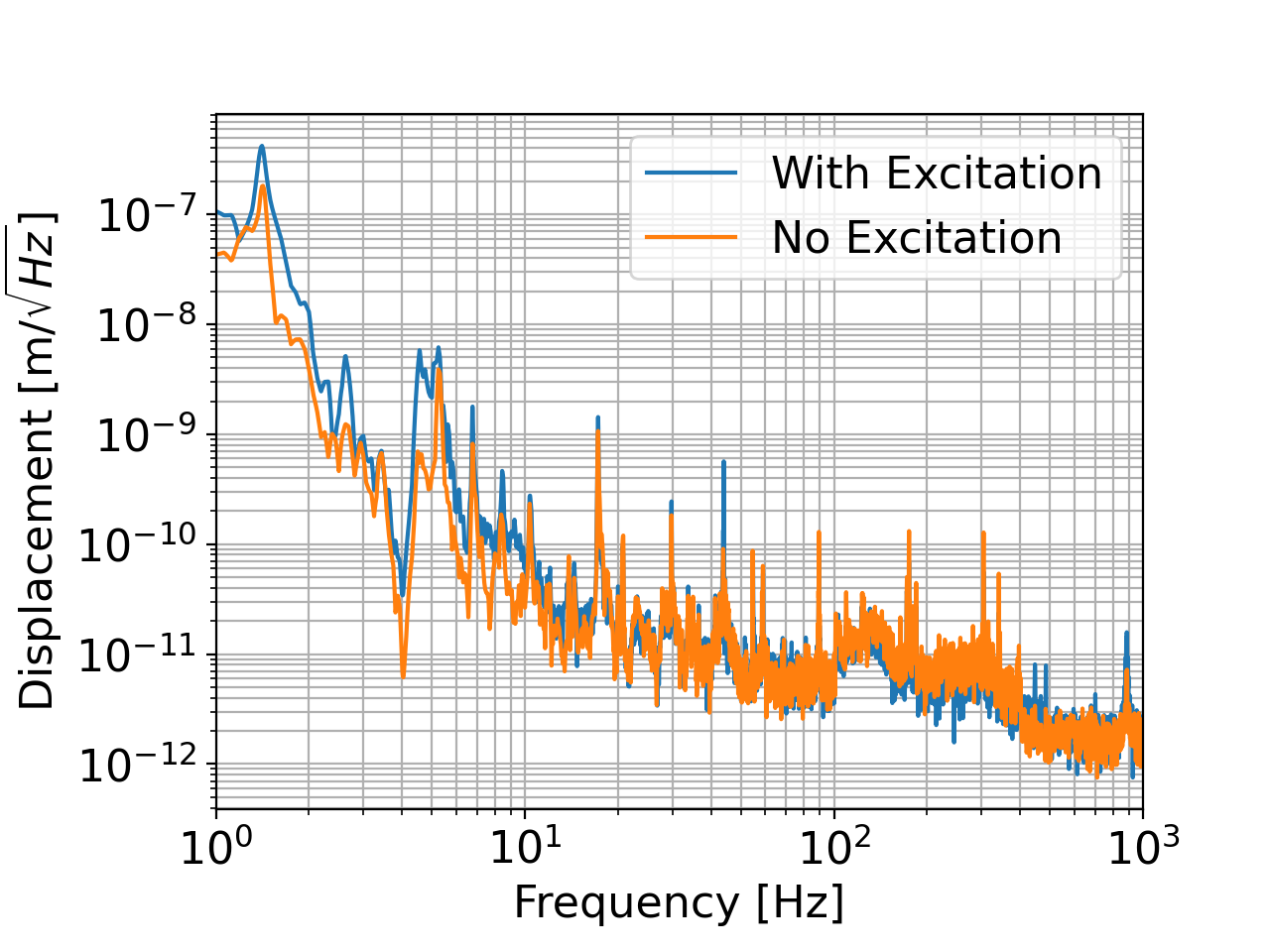}
\caption{Power spectrum of the vertical displacement in the CNMS. The orange curve is the power spectrum without excitation (0.1\,Hz), and the blue one is with excitation.}\label{sens}
\end{center}
\end{figure}

We conducted the experiment at the KAGRA site, where the seismic noise is generally small. To excite crackling noise in CNMS, a sinusoidal signal was applied with amplitude $1.7\times10^{-5}$\,m and frequency 0.1\,Hz to the coil-magnet actuator for the keystone. Figure~\ref{sens} is the comparison of the feedback signal spectrum of the CNMS control loop with and without excitation, which are calibrated to the differential displacement of mirrors. According to this figure, the noise floors of the two measurements are almost equal at broad frequencies above 20\,Hz. There is a noise increase observed below 20Hz, which can be crackling noise or induced by other unknown mechanisms. Although we are not confident that is crackling noise, an upper limit of crackling noise can be set.

\begin{figure}[htbp]
\begin{center}
\includegraphics[width=8cm]{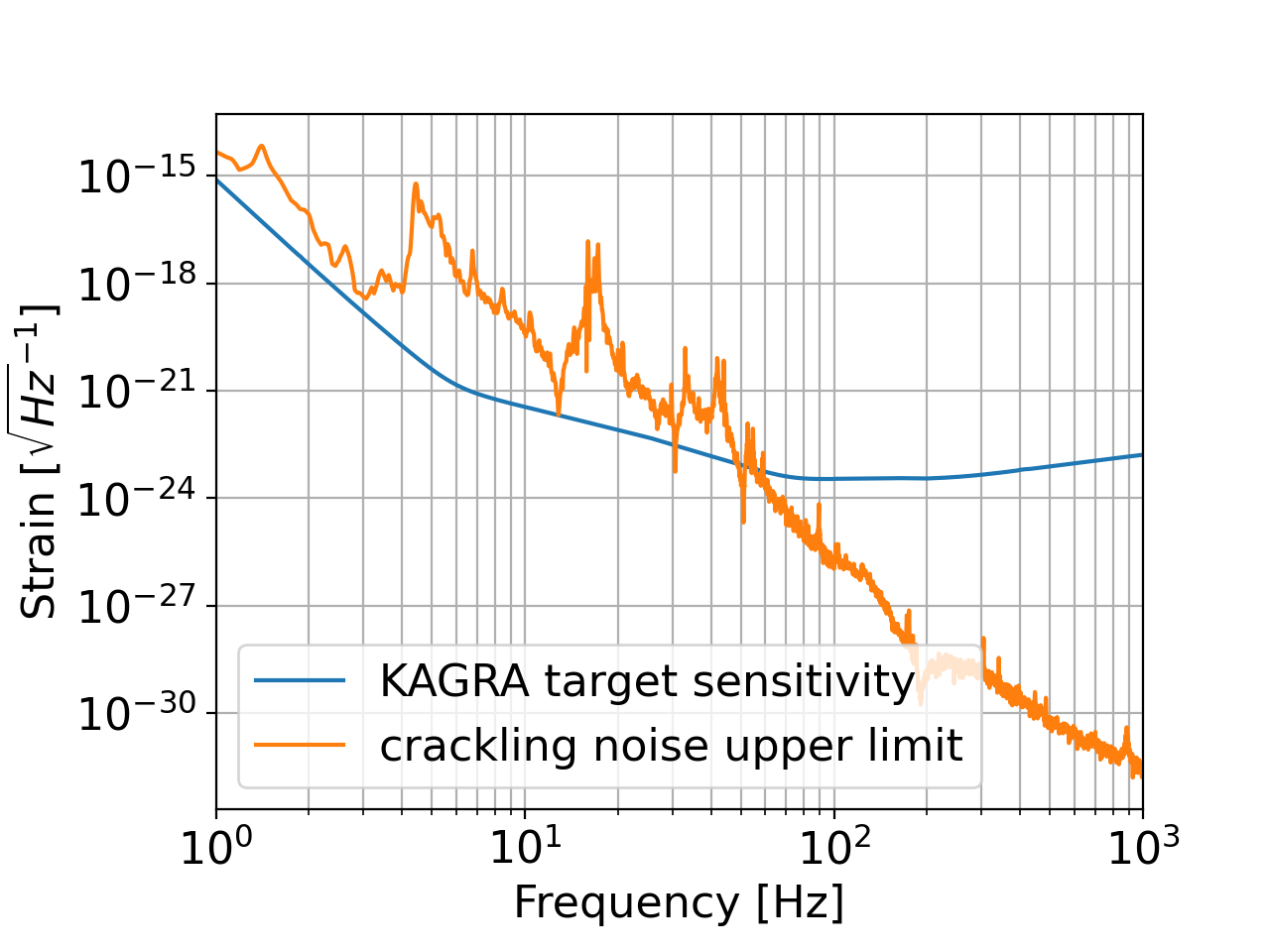}
\caption{Upper limit of the crackling noise effect in KAGRA together with the KAGRA target sensitivity.}\label{um_cra}
\end{center}
\end{figure}

By multiplying equation~\ref{convertfunction} with the displacement noise with excitation in figure~\ref{sens}, we can scale the crackling noise from CNMS to KAGRA type-A. To do this scaling, parameters in table~\ref{tab: sca} and excitation parameters for CNMS were used. The micro-seismic motion excites BF in KAGRA type-A, which has an amplitude of $5.7\times10^{-8}$\,m, and frequency of 0.1\,Hz. The difference of this oscillating load is considered in scaling CNMS result to Type-A, but the factor $\varepsilon$ is taken as $1$ to provide an upper limit. Figure~\ref{um_cra} shows the comparison of crackling noise upper limit and KAGRA design sensitivity.

\section{Conclusion and outlook}

Crackling noise is prevalent in many materials, which stems from the discrete and impulsive events when systems are exposed to varying external conditions. To investigate its impact on the KAGRA sensitivity, we set up a tailored experiment using the CNMS and tried to measure the crackling noise. We also derived a scaling law to convert the obtained result with the CNMS to estimate the crackling noise for KAGRA. Figure\,\ref{um_cra} shows that the upper limit of the crackling noise is lower than the KAGRA target sensitivity above 60\,Hz, whereas the upper limit is larger by up to a factor of 30 at frequencies below 60\,Hz.

 Here we should emphasize that the obtained curve in figure\,\ref{um_cra} is not the estimated crackling noise but the upper limit of the crackling noise for KAGRA. In the previous section, we explained that we did not observe any increase in the feedback spectrum with the excitation above 20 Hz. Thus above 20 Hz, the curve in figure\,\ref{um_cra} is the upper limit. Below 20\,Hz, although we observed a noise increase, we are not confident if the noise increase is caused by the crackling noise or by other unknown mechanisms. In this sense, even below 20\,Hz, the curve in figure\,\ref{um_cra} is considered to be an upper limit.
 
In addition, it should be noted that our crackling noise model does not consider the attenuation of the force release inside the GAS because we were not confident with any models. However, according to a simulation of GAS \cite{zhang}, the stress is maximized in the wide end of the GAS blade. If this is correct, KAGRA’s larger blade size could provide more considerable attenuation for the crackling noise. This could make our estimation of the crackling noise upper limit a conservative one.
 It is true that the upper-limit crackling noise is larger than the KAGRA sensitivity below 60\,Hz. To better examine the crackling noise, in the future, we should improve the sensitivity of the CNMS especially below 60\,Hz, to provide a better estimation of the crackling noise spectrum.
 
It was found in Ref.\,\cite{21_crack}, the onset of dislocation movement increases drastically with temperature increase. Although this was a measurement of creep noise, crackling noise shares the same origin of dislocation movement. The creep noise increases because higher temperature makes material tend to have more plastic deformation \cite{miguel}, which may induce more crackling noise and needs to be more investigated in the future.

\section*{Acknowledgment}
This work was supported by MEXT, JSPS Leading-edge Research Infrastructure Program, JSPS Grant-in-Aid for Specially Promoted Research 26000005, MEXT Grant-in-Aid for Scientific Research on Innovative Areas 24103005, JSPS Core-to-Core Program, A. Advanced Research Networks, the joint research program of the Institute for Cosmic Ray Research, University of Tokyo, the LIGO project, and the National Natural Science Foundation of China under Grants No.11633001. We would like to thank Yuji Yamanaka for his contributions to the early stage of the research, Gabriele Vajente, Xiaoyue Ni, and Rana Adhikari for their helpful advice, Mark Barton, Marc Eisenmann, and Michael Page for their suggestion on English.
\end{document}